\documentclass[a4paper,11pt]{article}

\usepackage{amsmath,amssymb,amsfonts,amsthm}
\usepackage[titletoc,title]{appendix}
\usepackage[bookmarks=true]{hyperref}

\numberwithin{equation}{section}

\newcommand{\bea}{\begin{eqnarray}}
\newcommand{\eea}{\end{eqnarray}}
\newcommand{\be}{\begin{equation}}
\newcommand{\ee}{\end{equation}}
\newcommand{\nn}{\nonumber \\}

\allowdisplaybreaks[1]

\begin{document}
\begin{titlepage}

\vspace*{0.2cm}

\begin{center}

{\LARGE\bf SU$(2|1)$ supersymmetric spinning models of}\\
\vspace{0.3cm}
{\LARGE\bf chiral superfields}

\vspace{1.5cm}

{\large\bf Stepan~Sidorov}
 \vspace{0.5cm}

{ \it Bogoliubov Laboratory of Theoretical Physics, JINR,}\\
{\it 141980 Dubna, Moscow region, Russia} \\

\vspace{0.1cm}

{\tt sidorovstepan88@gmail.com}\\
\vspace{0.7cm}

\end{center}
\vspace{0.2cm} \vskip 0.6truecm \nopagebreak

   \begin{abstract}
\noindent
We construct SU$(2|1)$, $d=1$ supersymmetric models based on the coupling of dynamical and semi-dynamical (spin) multiplets, where the interaction term of both multiplets is defined on the generalized chiral superspace. The dynamical multiplet is defined as a chiral multiplet ${\bf (2,4,2)}$, while the semi-dynamical multiplet is associated with a multiplet ${\bf (4,4,0)}$ of the mirror type.

\end{abstract}

\vspace{1.5cm}
\bigskip
\noindent PACS: 11.30.Pb, 12.60.Jv

\smallskip
\noindent Keywords: supersymmetric mechanics, deformation, spin variables

\newpage

\end{titlepage}
\section{Introduction}
A new class of systems of ${\cal N}=4$, $d=1$ supersymmetric quantum mechanics called ``K\"ahler oscillator'' was introduced by S. Bellucci and A. Nersessian in \cite{BelNer}. They studied supersymmetric oscillator models on K\"ahler manifolds with Wess-Zumino (WZ) type terms responsible for the presence of a constant magnetic field. It turned out, the presence of oscillator and WZ terms deforms the standard ${\cal N} = 4$ Poincar\'e supersymmetry to the so called ``weak supersymmetry'' \cite{WS}. In our works \cite{DSQM,SKO}, we showed that the deformed superalgebra of weak supersymmetry corresponds to SU$(2|1)$ supersymmetry. In these papers we initiated a study of deformed supersymmetric quantum mechanics by employing superfield approach based on the worldline supersymmetry SU$(2|1)$ with a mass dimension deformation parameter $m$. In the limit $m=0$, models of the standard ${\cal N}=4$ supersymmetric mechanics are restored. Indeed, within this framework we reproduced the models studied earlier at the component level \cite{BelNer,WS,Romelsberger} and obtained the new ones, including those constructed via harmonic superspace \cite{DHSS} (see also \cite{FI}).

During the study of SU$(2|1)$ supersymmetric mechanics we revealed few peculiar features. One of them is the generalization of a chiral condition given by \cite{SKO}
\bea
	\left(\cos{\lambda}\;\bar{{\cal D}}_i - \sin{\lambda}\;{\cal D}_i\right)\varphi = 0,\label{chiral0}
\eea
where ${\cal D}_i$ and $\bar{{\cal D}}_i$ are SU$(2|1)$ covariant derivatives. It describes a new type of the chiral multiplet ${\bf (2,4,2)}$ defined on the generalized chiral superspace and depending on two deformation parameters: $\lambda$ and $m$. Exactly this multiplet is a basis for the construction of supersymmetric K\"ahler oscillator models \cite{BelNer}, where the frequency of oscillator and the strength of external magnetic field are identified with
\bea
	 \omega = \frac{m\sin{2\lambda}}{2}\,,\qquad B=m\cos{2\lambda}\,.
\eea
However, both parameters disappear in the limit $m=0$, because the rotation parameter $\lambda$ becomes just an external automorphism parameter of the standard ${\cal N} = 4$ Poincar\'e supersymmetry. One needs to point out that the corresponding Hamiltonian does not commute with supercharges and is identified with internal U(1) generator of SU$(2|1)$. It can be treated as a central charge only when $\lambda=0$. Recently, particular superintegrable K\"ahler oscillator models were considered \cite{Superintegrable}.

With the lapse of time we found out other distinguished features of deformed supersymmetric quantum mechanics. An important one concerns ${\cal N}=4$ ``mirror multiplets'' \cite{Niederle}. The standard ${\cal N}=4$ multiplets have their mirror counterparts characterized by the interchange of two SU$(2)$ groups which form SO$(4)$ automorphism group of the standard ${\cal N}=4$ Poincar\'e supersymmetry. Since this interchange has no essential impact on Poincar\'e supersymmetry, ${\cal N}=4$ multiplets and their mirror counterparts are mutually equivalent when dealing with only one multiplet from such a pair. Deformation to SU$(2|1)$ supersymmetry breaks this equivalence, because the first SU$(2)$ group becomes subgroup of SU$(2|1)$ and the second one is broken. It means that SU$(2|1)$ multiplets differ from their mirror counterparts. We showed this difference in details on the example of ${\bf (4,4,0)}$ multiplets \cite{DHSS}.

The main goal of the present paper is to employ SU$(2|1)$ superfield approach to spinning models of chiral superfields instead of harmonic ones. The SU$(2|1)$ supersymmetric spinning models of harmonic superfields \cite{FI, FIS, FILS} followed the construction elaborated in \cite{FIL1, FIL2}, with the dynamical multiplet ${\bf (1,4,3)}$ and the semi-dynamical multiplet ${\bf (4,4,0)}$. The most important property is that both dynamical and semi-dynamical multiplets admit a description in the analytic harmonic superspace \cite{IL}, like as the ${\cal N}=4$, $d=1$ gauge multiplet \cite{DI}. The latter one allows to introduce the U$(1)$ gauge symmetry transformations. The gauge invariant coupling term is constructed as an integral over the analytic harmonic superspace and provides the interaction of two multiplets. This term is identified with WZ term and involves semi-dynamical spin variables \footnote{Some recent ${\cal N}=4$ deformed and undeformed supersymmetric models with spin variables were studied at the component level in \cite{KKLSKLSGL}.}. Extra SU(2) symmetries are defined in terms of these variables, with respect to which physical states of quantum models carry additional spin quantum numbers. Generalization of the construction to matrix superfields \cite{FIL1} give rise to ${\cal N}=4$ supersymmetric extentions of the widely known Calogero system \cite{Polychronakos1} (see also the review \cite{Polychronakos2}).

In this paper we consider the coupling of the dynamical multiplet ${\bf (2,4,2)}$ and the semi-dynamical multiplet ${\bf (4,4,0)}$ of the mirror type. Both multiplets satisfy the generalized chiral condition \eqref{chiral0} that allows to introduce the interaction term as a superpotential term. We construct SU$(2|1)$ supersymmetric models based on this coupling, undeformed versions of which were studied in \cite{BKKS} \footnote{In this paper, superfield description for a semi-dynamical multiplet proceeds from the fermionic multiplet ${\bf (0,4,4)}$. One can make the same comment as in the footnote 7 of \cite{I}.}. To compare the results with \cite{BKKS}, we consider an example of a model on the pseudo-sphere SU$(1,1)$/U$(1)$ (the Lobachevsky space).

\section{SU(2$|$1) chiral superfields}
In \cite{SKO}, we considered a special type of SU$(2|1)$ superspace defined for the standard form of the superalgebra $su(2|1)$:
\bea
    &&\lbrace Q^{i}, \bar{Q}_{j}\rbrace = 2m I^i_j +2\delta^i_j {\cal H},\qquad
    \left[I^i_j,  I^k_l\right]
    = \delta^k_j I^i_l - \delta^i_l I^k_j\,,\nn
    &&\left[I^i_j, \bar{Q}_{l}\right] = \frac{1}{2}\,\delta^i_j\bar{Q}_{l}-\delta^i_l\bar{Q}_{j}\, ,\qquad \left[I^i_j, Q^{k}\right]
    = \delta^k_j Q^{i} - \frac{1}{2}\,\delta^i_j Q^{k},\nn
    &&\left[{\cal H}, \bar{Q}_{l}\right]=\frac{m}{2}\,\bar{Q}_{l}\,,\qquad \left[{\cal H}, Q^{k}\right]=-\frac{m}{2}\,Q^{k}. \label{algebra1}
\eea
The bosonic subgroup includes of SU$(2)$ subgroup composed of $I^i_j$ ($i=1, 2$) and U(1) generator ${\cal H}$ associated with Hamiltonian. Factorizing the SU$(2)$ subgroup, the superspace is defined as
\bea
	\frac{{\rm SU}(2|1)}{{\rm SU}(2)}\,\sim\,\frac{\left\lbrace{\cal H}, Q^{i}, \bar{Q}_{j}\,, I^i_j\right\rbrace}{\left\lbrace I^i_j\right\rbrace}=\left\lbrace t, \theta_i, \bar{\theta}^j\right\rbrace.
\eea
The odd $\epsilon^i$ transformations are given by
\bea
    &&\delta t=i\left(\epsilon_k\bar{\theta}^k+\bar{\epsilon}^k\theta_k\right),\qquad
    \delta\theta_{i}=\epsilon_{i}+2m\,\bar{\epsilon}^k\theta_k \theta_{i}\,,\qquad\delta\bar{\theta}^{j}
    =\bar{\epsilon}^{i}-2m\,\epsilon_k\bar{\theta}^k\bar{\theta}^{i},\nn
    &&\left(\theta_i\right)=\bar{\theta}^i,\qquad\left(\epsilon_i\right)=\bar{\epsilon}^i. \label{SU21_tr}
\eea
In the limit $m=0$, the standard ``flat'' supersymmetry transformations are restored. One can easily pick out the chiral subspace $\left\lbrace t_{\rm L},\theta_i\right\rbrace$, where
\bea
	t_{\rm L} = t +i\,\bar{\theta}^k\theta_k - im\left(\bar{\theta}^k\theta_k\right)^2,\qquad\delta t_{\rm L}=2i\,\bar{\epsilon}^k\theta_k\,,\qquad
    \delta\theta_{i}=\epsilon_{i}+2m\,\bar{\epsilon}^k\theta_k \theta_{i}\,.
\eea
We also define the generalized chiral coordinates
\bea
	\hat{t}_{\rm L} = t +i\,\bar{\hat\theta}^k\hat{\theta}_k \,,\qquad
    \hat{\theta}_i = \left(\cos{\lambda}\,\theta_i\, e^{\frac{i}{2}m t}+\sin{\lambda}\,
    \bar{\theta}_i\,e^{-\frac{i}{2}m t}\right)
    \left(1 -\frac{m}{2}\,\bar{\theta}^k\theta_k\right).\label{left}
\eea
They are closed under the following supersymmetry transformations:
\bea
    &&\delta\hat{\theta}_{i}=\cos{\lambda}\left(\epsilon_{i}\,e^{\frac{i}{2}m \hat{t}_{\rm L}}
    + m\,\bar{\epsilon}^k\hat{\theta}_k\hat{\theta}_{i}\, e^{-\frac{i}{2}m\hat{t}_{\rm L}}\right)
    +\sin{\lambda}\left(\bar{\epsilon}_i\,e^{-\frac{i}{2}m\hat{t}_L} + m\,\epsilon^k\hat{\theta}_k\hat{\theta}_{i}\, e^{\frac{i}{2}m\hat{t}_{\rm L}}\right),\nn
    && \delta \hat{t}_{\rm L}=2i\left(\cos{\lambda}\;\bar{\epsilon}^k\hat{\theta}_k\, e^{-\frac{i}{2}m\hat{t}_{\rm L}} -\sin{\lambda}\;\epsilon^k\hat{\theta}_k \,e^{\frac{i}{2}m\hat{t}_{\rm L}}\right).\label{left_tr} 
\eea
Chiral superfields, defined on the left chiral subspace $\left\lbrace\hat{t}_{\rm L},\hat\theta_i\right\rbrace$, satisfy the chiral condition
\bea
	{\bar{\tilde{\cal D}}}_j\,\varphi\left(\hat{t}_{\rm L}, \hat{\theta}\right) = 0,\label{chiral}
\eea
where the covariant derivatives ${\bar{\tilde{\cal D}}}_i$ and ${\tilde{\cal D}}^i$ are modified as
\bea
	{\bar{\tilde{\cal D}}}_i=\cos{\lambda}\;\bar{{\cal D}}_i -\sin{\lambda}\;{\cal D}_i\,,\qquad {\tilde{\cal D}}^i=\cos{\lambda}\;{\cal D}^i +\sin{\lambda}\;\bar{{\cal D}}^i\,,\qquad \lambda \in \left[0,\pi/2\right).
\eea
In the basis $\left\lbrace\hat{t}_{\rm L},\hat{\theta}_j,\bar{\hat\theta}^i\right\rbrace$\,, the covariant derivative ${\bar{\tilde{\cal D}}}_i$ is written as
\bea
    \bar{\tilde{{\cal D}}}_j = -\left[1+\frac{m\cos{2\lambda}}{2}\,\bar{\hat\theta}^k\hat{\theta}_k
    - \frac{m\sin{2\lambda}}{4}\left(\bar{\hat\theta}^k\bar{\hat\theta}_k+\hat{\theta}_k\hat{\theta}^k\right)-\frac{m^2}{16}\,\hat{\theta}_i\hat{\theta}^i\,\bar{\hat{\theta}}^j\bar{\hat{\theta}}_j\right]\frac{\partial}{\partial\bar{\hat{\theta}}^j}\,.
\eea
Here, we ignore the matrix SU(2) generators $\tilde{I}^i_j$\,, because the generalized chiral superfields ${\bar{\tilde{\cal D}}}_i$
cannot carry any external SU(2) indices \footnote{The explanation is given in Appendix B of \cite{IST}.\label{f3}}.

The SU$(2|1)$ invariant measure is written as
\bea
    d\zeta = dt\,d^2\theta\,d^2\bar{\theta}\left(1+ 2m\,\bar{\theta}^k\theta_k
    \right).\label{measure}
\eea
One can write this measure in the basis $\left\lbrace t,\hat\theta_i,\bar{\hat\theta}^j\right\rbrace$ as
\bea
	d\zeta = dt\, d^2\hat\theta\, d^2\bar{\hat{\theta}}\left[1+ m\cos{2\lambda}\;\bar{\hat{\theta}}^k\hat{\theta}_k
    -\frac{m\sin{2\lambda}}{2}\left(\hat{\theta}_k\hat{\theta}^k +\bar{\hat{\theta}}^k\bar{\hat{\theta}}_k\right)\right].\label{measure1}
\eea
The chiral measure $d\hat{t}_{\rm L}\,d^2\hat{\theta}$ is also invariant under the transformations \eqref{left_tr}:
\bea
	d\zeta_{\rm L}=d\hat{t}_{\rm L}\,d^2\hat{\theta},\qquad \delta\left(d\zeta_{\rm L}\right)=0.
\eea
\subsection{Dynamical multiplet (2,4,2)}
The multiplet ${\bf (2,4,2)}$ is described by a complex superfield $\Phi$ subjected to the chirality condition \cite{SKO}
\bea
   {\bar{\tilde{\cal D}}}_j\Phi = 0.
\eea
Its general solution reads
\bea
	\Phi\left(\hat{t}_{\rm L}, \hat{\theta}\right) = z+\sqrt{2}\,\hat{\theta}_k\xi^k + \hat{\theta}_k\hat{\theta}^k B.
\eea
The superfield is not deformed, but transformation properties of its components are still deformed:
\bea
    &&\delta z  =-\,\sqrt{2}\cos{\lambda}\;\epsilon_k\xi^k\,e^{\frac{i}{2}m t} - \sqrt{2}\sin{\lambda}\;\bar{\epsilon}_k\xi^k\,e^{-\frac{i}{2}mt},\nn
    &&\delta \xi^k =\sqrt{2}\,\bar{\epsilon}^k\left(i\cos{\lambda}\;\dot z -\sin{\lambda}\;B\right)e^{-\frac{i}{2}m t}-\sqrt{2}\,\epsilon^k\left(i\sin{\lambda}\;\dot z + \cos{\lambda}\;B\right) e^{\frac{i}{2}m t},\nn
    &&\delta B = -\,\sqrt{2}\cos{\lambda}\;\bar{\epsilon}_k\left(i\,\dot{\xi}^k+\frac{m}{2}\,\xi^k\right)e^{-\frac{i}{2}m t}+\sqrt{2}\sin{\lambda}\;\epsilon_k\left(i\,\dot{\xi}^k-\frac{m}{2}\,\xi^k\right) e^{\frac{i}{2}m t}.
\eea
The deformation parameters $m$ and $\lambda$ also appear in \eqref{measure1}. 

Superfield kinetic action for the chiral superfield $\Phi$ is given by
\bea
	S_{\rm kin.}=\frac{1}{4}\int d\zeta\,K\left(\Phi,\bar{\Phi}\right).\label{kinetic}
\eea
where $K\left(\Phi,\bar{\Phi}\right)$ is a K\"ahler potential.
The component Lagrangian is written as
\bea
	{\cal L}_{\rm kin.}&=&g\dot{\bar{z}}\dot{z}
    + \frac{i}{2}\,g\left(\xi^i\dot{\bar{\xi}}_i-\dot{\xi}^i\bar{\xi}_i\right) + g\,B\bar{B}-\frac{B}{2}\,\partial_{\bar z}g\,\bar{\xi}^k\bar{\xi}_k -\frac{\bar{B}}{2}\,\partial_{z}g\,\xi_k\xi^k\nn
    &&+\,\frac{i}{2}\left(\dot{\bar{z}}\,\partial_{\bar z} g -\dot{z}\,\partial_{z}g\right)\xi^k\bar{\xi}_k  +\frac{1}{4}\,\partial_{z}\partial_{\bar z}g\,\xi_i\xi^i\,\bar{\xi}^j\bar{\xi}_j-\frac{i}{2}\,m\cos{2\lambda}\left(\dot{\bar{z}}\,\partial_{\bar z} K - \dot{z}\,\partial_{z} K\right)\nn
    &&-\,\frac{m\cos{2\lambda}}{2}\,g\,\xi^k\bar{\xi}_k-\frac{m\sin{2\lambda}}{2}\left(\bar{B}\,\partial_{\bar z}K+B\,\partial_z K\right)\nn
    &&+\,\frac{m\sin{2\lambda}}{4}\left(\partial_z\partial_z K\,\xi_i\xi^i+\partial_{\bar z}\partial_{\bar z} K\,\bar{\xi}^j\bar{\xi}_j\right).\label{kin.}
\eea
with the metric $g\left(z,\bar{z}\right):=\partial_{z}\partial_{\bar z}K\left(z,\bar{z}\right)$\,.

\subsection{Semi-dynamical mirror multiplet (4,4,0)}
In \cite{DHSS}, we studied the deformed mirror multiplet ${\bf (4, 4, 0)}$ in the framework of harmonic superspace. Here we consider its generalization to the chiral superspace \eqref{left}. The generalized mirror multiplet ${\bf (4, 4, 0)}$ is described by a pair of chiral superfields $Y^A$ ($A=1,2$) satisfying the constraints
\bea
    {\bar{\tilde{\cal D}}}_i Y^{A}=0,\qquad {\tilde{\cal D}}^i\bar{Y}^{A} = 0,\qquad {\tilde{\cal D}}_i Y^{A}={\bar{\tilde{\cal D}}}_i\bar{Y}^{A},\qquad
    \overline{\left(Y^{A}\right)}=\bar{Y}_{A}\,.\label{440constr}
\eea
Their solution reads
\bea
    Y^A\left(\hat{t}_{\rm L}, \hat{\theta}\right)=y^{A} + \sqrt{2}\,\hat{\theta}_{i} \psi^{iA} + i\,\hat{\theta}_{k}\hat{\theta}^{k}\,\dot{\bar{y}}^{A},\label{Y_L}
\eea
where
\be
    \overline{\left(y^{A}\right)} = \bar{y}_{A}\,,
\qquad \overline{\left(\psi^{iA}\right)} = \psi_{iA}\,.
\ee
The corresponding component field transformations are
\bea
    &&\delta y^A  =-\,\sqrt{2}\cos{\lambda}\;\epsilon_k\psi^{kA}\,e^{\frac{i}{2}m t} - \sqrt{2}\sin{\lambda}\;\bar{\epsilon}_k\psi^{kA}\,e^{-\frac{i}{2}mt},\nn
    &&\delta \bar{y}_A  =\sqrt{2}\cos{\lambda}\;\bar{\epsilon}^k\psi_{kA}\,e^{-\frac{i}{2}m t} - \sqrt{2}\sin{\lambda}\;\epsilon^k\psi_{kA}\,e^{\frac{i}{2}mt},\nn    
    &&\delta \psi^{kA} =\sqrt{2}\,i\,\bar{\epsilon}^k\left(\cos{\lambda}\;\dot{y}^A - \sin{\lambda}\;\dot{\bar{y}}^A\right)e^{-\frac{i}{2}m t}-\sqrt{2}\,i\,\epsilon^k\left(\cos{\lambda}\;\dot{\bar{y}}^A+\sin{\lambda}\;\dot{y}^A\right) e^{\frac{i}{2}m t}.\nn\label{trY}
\eea

The invariant superfield action is written as
\bea
    S_{\bf (4,4,0)} =\frac{1}{2}\int d\zeta\,L\left(Y,\bar{Y}\right).\label{Gaction-Y}
\eea
The component Lagrangian contains the standard kinetic term $G\,\dot{y}^{A}\dot{\bar{y}}_{A}$, where the metric $G$ of 4-dimensional manifold is defined as
\bea
    G\left(y,\bar{y}\right):=\Delta_y L\left(y,\bar{y}\right) ,\qquad \Delta_y  = -\,2\,\varepsilon^{AB}\partial_{A}\bar{\partial}_{B}\,,\qquad \partial_A
    = \frac{\partial}{\partial y^A}\,,\quad
    \bar{\partial}_B = \frac{\partial}{\partial \bar{y}^B}\,.
\eea
In order to describe the semi-dynamical spin multiplet we must drop such kinetic terms. So, we take the limit $G=0$ that gives the residual Lagrangian of the first order in time derivatives (Wess-Zumino type Lagrangians) where the function $L(y,\bar{y})$ satisfies the Laplace equation \be \Delta_y L=0. \ee
Another option is given by the superpotential term
\bea
	S_{\rm pot.}=\frac{\mu}{2}\int d\zeta_{\rm L}\,h\left(Y^A\right) + \frac{\mu}{2}\int d\zeta_{\rm R}\,h\left(\bar{Y}_A\right).\label{spot}
\eea
One can consider the function $h\left(y^A\right)$ in \eqref{hYY} as a holomorphic subset of the solutions found in \cite{DHSS}, where the general construction of Wess-Zumino type Lagrangians was given.

Here, we are interested in coupling of this semi-dynamical multiplet with the chiral superfield $\Phi$ that can be described only by a superpotential term. So, we skip the construction of Wess-Zumino type Lagrangians from \eqref{Gaction-Y} and present the superpotential term written as
\bea
	S_{\rm pot.}=\int dt\,{\cal L}_{\rm WZ}\,,\qquad {\cal L}_{\rm WZ} = \mu\left[i\,\dot{\bar{y}}^{A}\,\partial_A h\left(y^A\right)+\frac{1}{2}\,\psi^{iA}\psi^{B}_{i}\,\partial_A\partial_B h\left(y^A\right)+{\rm c.c.}\right]\,.\label{hYY}
\eea
This Lagrangian contains no deformation parameters, so it is invariant as well under the standard ($m=0$) supersymmetry transformations. Indeed, the superpotential term must be invariant under the transformations closed on the centerless ${\cal N} = 4$ super Virasoro algebra \cite{HT}, which is an infinite dimensional superalgebra possessing both deformed and undeformed superalgebras.

\subsection{Gauged mirror multiplet (4,4,0)}
By analogy with \cite{FI}, we can also expect that chiral superfields are subjected to the local U$(1)$ transformations
\bea
	&&\left(Y^A\right)^{\prime} = e^{\left(\Lambda - \bar{\Lambda}\right){\cal I}}\,Y^{A},\qquad {\cal I}\,Y^{A}=\frac{1}{2}\left(\sigma_3\right)^A_B Y^{B},\nn
	&&\left(\bar{Y}^A\right)^{\prime} = e^{\left(\Lambda - \bar{\Lambda}\right){\cal I}}\,\bar{Y}^{A},\qquad {\cal I}\,\bar{Y}^{A}=\frac{1}{2}\left(\sigma_3\right)^A_B \bar{Y}^{B},\label{YU1}
\eea
where ${\cal I}$ is a U$(1)$ generator and $\Lambda=\Lambda\left(\hat{t}_{\rm L},\hat{\theta}\right)$, $\bar{\Lambda}=\bar{\Lambda}\left(\hat{t}_{\rm R},\bar{\hat{\theta}}\,\right)$.
Gauged superfields satisfy the new gauge invariant constraints
\bea
	&&\left({\bar{\tilde{\cal D}}}^i + \left[{\bar{\tilde{\cal D}}}^i,X\right]{\cal I}\right)Y^{A}=0,\qquad \left(\tilde{\cal D}^i - \left[\tilde{\cal D}^i, X\right]{\cal I}\right)\bar{Y}^{A} = 0,\label{1}\\
	&&\left(\tilde{\cal D}^i - \left[\tilde{\cal D}^i, X\right]{\cal I}\right)Y^{A}=\left({\bar{\tilde{\cal D}}}^i + \left[{\bar{\tilde{\cal D}}}^i,X\right]{\cal I}\right)\bar{Y}^{A},\label{2}
\eea
where the real superfield $X$ is a gauge superfield transforming as
\bea
	X^\prime = X + \Lambda + \bar{\Lambda}\,.\label{XLambda}
\eea
Superspace expansion of the gauge superfield $X$ displays 8 bosonic and 8 fermionic components:
\bea
	X\left(t, \hat{\theta}, \bar{\hat{\theta}}\,\right)&=&x+\hat{\theta}_k\chi^k - \bar{\hat{\theta}}^k\bar{\chi}_k + 2\,\bar{\hat{\theta}}^k\hat{\theta}_k{\cal A}+ \hat{\theta}_k\hat{\theta}^k\,D + \bar{\hat{\theta}}^k\bar{\hat{\theta}}_k\,\bar{D}+\bar{\hat{\theta}}_{(i}\hat{\theta}_{j)}\,{\cal B}^{ij}\nn
	&&+\,\bar{\hat{\theta}}^k\hat{\theta}_k\left(\hat{\theta}_k\zeta^k - \bar{\hat{\theta}}^k\bar{\zeta}_k\right)+\hat{\theta}_i\hat{\theta}^i\,\bar{\hat{\theta}}^j\bar{\hat{\theta}}_j\,C,\label{8+8}
\eea
where
\bea
&& \overline{\left(x\right)} = x,\qquad\overline{\left({\cal A}\right)} = {\cal A}\,,\qquad \overline{\left(D\right)} = \bar{D},\qquad \overline{\left({\cal B}^{ij}\right)}=-\,{\cal B}_{ij},\qquad {\cal B}^{ij}={\cal B}^{ji}\,,\nn
&& \overline{\left(\chi^i\right)} = \bar{\chi}_i\,,\qquad
\overline{\left(\zeta^i\right)} = \bar{\zeta}_i\,.
\eea
However, from the constraints \eqref{1} and \eqref{2} we obtain the additional constraint \footnote{Actually, the gauge superfield $X$ can be considered as a prepotential for the vector multiplet ${\bf (3,4,1)}$ given by the superfield $V_{ij}=\tilde{\cal D}_{(i}{\bar{\tilde{\cal D}}}_{j)}X$ \cite{IvSm} satisfying the constraints
$$ {\bar{\tilde{\cal D}}}_{(i}V_{jk)}=\tilde{\cal D}_{(i}V_{jk)}=0.$$
One can check that it is invariant under the gauge transformations \eqref{XLambda}. Hence, the following constraint \eqref{143c} implies that all physical degrees of freedom of \eqref{8+8} can be gauged away, {\it i.e.} the gauge multiplet becomes ``topological'' \cite{DI}.}
\bea
	\tilde{\cal D}_{(i}{\bar{\tilde{\cal D}}}_{j)}X=0.\label{143c}
\eea
It kills half of the components of \eqref{8+8} as
\bea
	X\left(t, \hat{\theta}, \bar{\hat{\theta}}\,\right)&=&x+\hat{\theta}_k\chi^k - \bar{\hat{\theta}}^k\bar{\chi}_k+2\,\bar{\hat{\theta}}^k\hat{\theta}_k{\cal A}+ \hat{\theta}_k\hat{\theta}^k\,D + \bar{\hat{\theta}}^k\bar{\hat{\theta}}_k\,\bar{D}\nn
	&&+\,i\,\bar{\hat{\theta}}^k\hat{\theta}_k\left(\hat{\theta}_i\dot{\chi}^i + \bar{\hat{\theta}}^i\dot{\bar{\chi}}_i\right)-\frac{1}{4}\,\hat{\theta}_i\hat{\theta}^i\,\bar{\hat{\theta}}^j\bar{\hat{\theta}}_j\,\ddot{x}.\label{143}
\eea
This superfield describes the mirror multiplet ${\bf (1,4,3)}$ that differs from the ordinary multiplet ${\bf (1,4,3)}$ \cite{DSQM} because of the deformation. Using the U$(1)$ gauge freedom \eqref{XLambda}, we can choose the WZ gauge 
\bea
	X_{\rm WZ}=2\,\bar{\hat{\theta}}^k\hat{\theta}_k\,{\cal A}\,,\qquad {\cal A}^{\prime} = {\cal A}-\dot{\alpha}\,.\label{XWZ}
\eea
Thus, it can be interpreted as a mirror counterpart of the ``topological'' gauge multiplet \cite{DI} described by the harmonic superfield $V^{++}$ in the WZ gauge. One can introduce accompanying chiral superfields
\bea
	{\cal V}_{\rm WZ}\left(\hat{t}_{\rm L}, \hat{\theta}\right)=\hat{\theta}_k\hat{\theta}^k{\cal A}\,,\qquad \bar{\cal V}_{\rm WZ}\left(\hat{t}_{\rm R}, \bar{\hat{\theta}}\,\right)=\bar{\hat{\theta}}^k\bar{\hat{\theta}}_k\,{\cal A}\,,\label{VWZ}
\eea
satisfying
\bea
	{\tilde{\cal D}}_i\,X_{\rm WZ} ={\bar{\tilde{\cal D}}}_i\bar{\cal V}_{\rm WZ}\,,\qquad {\bar{\tilde{\cal D}}}_i\,X_{\rm WZ} =-\,{\tilde{\cal D}}_i{\cal V}_{\rm WZ}\,.\label{341m}
\eea
The triplet of superfields, given by \eqref{XWZ} and \eqref{VWZ}, has a harmonic superfield description in the flat superspace $m=0$ with respect to the second ${\rm SU}^{\prime}(2)$ subgroup \cite{Niederle}. One could ascribe ${\rm SU}^{\prime}(2)$ indices to this triplet as ${\cal V}^{(i^{\prime}j^{\prime})}$ and harmonize them. In the deformed supersymmetric mechanics we have no such description, because the second ${\rm SU}^{\prime}(2)$ symmetry is broken.

According to the chiral condition \eqref{1}, the superfield solution \eqref{Y_L} is modified as
\bea
    Y^A\left(t, \hat{\theta}, \bar{\hat{\theta}}\,\right)=e^{-X\,{\cal I}}\,Y^A_{\rm L}\left(\hat{t}_{\rm L},\hat{\theta}\right),\qquad
    \left(Y^A_{\rm L}\right)^{\prime} = e^{2\Lambda\,{\cal I}}\,Y^{A}_{\rm L}\,,\nn
    \bar{Y}^A\left(t, \hat{\theta}, \bar{\hat{\theta}}\,\right)=e^{X\,{\cal I}}\,\bar{Y}^A_{\rm R}\left(\hat{t}_{\rm R},\bar{\hat{\theta}}\,\right),\qquad
    \left(\bar{Y}^A_{\rm R}\right)^{\prime} = e^{-2\bar{\Lambda}\,{\cal I}}\,\bar{Y}^{A}_{\rm R}\,.\label{YU1L}
\eea
Solving \eqref{2}, the left chiral superfield $Y^A_{\rm L}$ has the $\theta$-expansion
\bea
    Y^A_{\rm L}\left(\hat{t}_{\rm L}, \hat{\theta}\right)=y^{A} + \sqrt{2}\,\hat{\theta}_{i} \psi^{iA} + i\,\hat{\theta}_{k}\hat{\theta}^{k}\,\nabla_t{\bar{y}}^{A},\label{gY_L}
\eea
where
\bea
	\nabla_t = \partial_t + 2i{\cal A}\,{\cal I}.
\eea
It is necessary to replace time derivatives by $\nabla_t$ in the transformations \eqref{trY}. The residual local $\alpha=\alpha\left(t\right)$ transformations of component fields must be written as
\bea
	\left(y^{A}\right)^{\prime} = e^{2i\alpha\,{\cal I}}\,y^{A},\quad
	\left(\psi^{iA}\right)^{\prime} = e^{2i\alpha\,{\cal I}}\,\psi^{iA},\quad
	\left(\nabla_t\bar{y}^{A}\right)^{\prime} = e^{2i\alpha\,{\cal I}}\,\nabla_t\bar{y}^{A},\qquad
	{\cal A}^{\prime} = {\cal A}-\dot{\alpha}\,.\label{tr_local}
\eea
Taking into account these transformations and the supersymmetry transformations \eqref{left_tr}, one can check that the real superfield \eqref{XWZ} transforms as \eqref{XLambda} with the parameter
\bea
	\Lambda\left(\hat{t}_{\rm L},\hat{\theta}\right)=i\alpha\left(\hat{t}_{\rm L}\right) + 2{\cal A}\left(\hat{t}_{\rm L}\right)\left(\cos{\lambda}\;\bar{\epsilon}^k\hat{\theta}_k\,e^{-\frac{i}{2}m\hat{t}_{\rm L}}-\sin{\lambda}\;\epsilon^k\hat{\theta}_k \,e^{\frac{i}{2}m\hat{t}_{\rm L}}\right).
\eea
The superfield $Y^A$ transforms with the same parameter according to \eqref{YU1}.

Finally, the superpotential \eqref{spot} must be written in terms of $Y^1Y^2\equiv Y^1_{\rm L}Y^2_{\rm L}$\,, since it is the only gauge invariant object defined on the left chiral subspace. One can define also the invariant F-term
\bea
	S_{\rm FI}=-\,\frac{c}{4}\left[\int d\zeta_{\rm L}\,{\cal V}_{\rm WZ} + \int d\zeta_{\rm R}\,\bar{\cal V}_{\rm WZ}\right]\quad\Rightarrow\quad {\cal L}_{\rm FI} = -\,c\,{\cal A}\,,\qquad c={\rm const},\label{FI}
\eea
which is a counterpart of Fayet-Iliopoulos term defined on the analytic harmonic superspace \cite{DI}.
\section{Coupling of dynamical and semi-dynamical multiplets}
In this section we consider the coupling of dynamical and semi-dynamical multiplets identified with the chiral multiplet ${\bf (2,4,2)}$ and the gauged mirror multiplet ${\bf (4,4,0)}$, respectively.

\subsection{Interacting superpotential term}
According to \eqref{YU1}, the simplest interacting superpotential term reads
\bea
	S_{\rm int.}=\frac{\mu}{2}\int d\zeta_{\rm L}\,Y^1Y^2\,f\left(\Phi\right) +\frac{\mu}{2}\int d\zeta_{\rm R}\,\bar{Y}_{1}\bar{Y}_{2}\,\bar{f}\left(\bar{\Phi}\right),\label{intAction}
\eea
where $f$ is an arbitrary holomorphic function of $\Phi$. 
The Lagrangian is then given by
\bea
	{\cal L}_{\rm int.} &=& \mu\,\bigg[i\left(y^1\,\nabla_t\bar{y}_{1}-y^2\,\nabla_t\bar{y}_{2}\right)f+\psi^{i1}\psi^{2}_{i}\,f  + B\,y^1 y^2\,\partial_z f\nn
	&&+\,\xi^{i}\left(\psi_{i1}\,y^1-\psi_{i2}\,y^2\right)\partial_z f
	-\frac{\xi_i\xi^i}{2}\,y^1 y^2\,\partial_z\partial_z f+{\rm c.c.}\,\bigg]\,.\label{int.}
\eea
It is straightforward to check that the Lagrangian is invariant under the local U$(1)$ transformations \eqref{tr_local}.

\subsection{Total Lagrangian}
The total Lagrangian of the interacting model is given by the sum of the kinetic Lagrangian \eqref{kin.}, the interacting superpotential term \eqref{int.} and Fayet-Iliopoulos term \eqref{FI}:
\bea
	{\cal L}={\cal L}_{\rm kin.}+{\cal L}_{\rm int.}+{\cal L}_{\rm FI}\,.
\eea
Eliminating the auxiliary fields $B$ and $\psi^{iA}$ by their equations of motion and performing the following redefinition
\bea
	&&y^1 = v\left(f+\bar{f}\,\right)^{-\frac{1}{2}},\qquad \bar{y}_1 = \bar{v}\left(f+\bar{f}\,\right)^{-\frac{1}{2}},\nn
	&&y^2 = \bar{w}\left(f+\bar{f}\,\right)^{-\frac{1}{2}},\qquad \bar{y}_2 = w\left(f+\bar{f}\,\right)^{-\frac{1}{2}},\nn
	&&\xi^i = g^{-\frac{1}{2}}\,\eta^i,\qquad \bar{\xi}_j = g^{-\frac{1}{2}}\,\bar{\eta}_j\,,
\eea
we obtain the total on-shell Lagrangian (up to full time derivatives) as
\bea
	{\cal L} &=& g\,\dot{\bar{z}}\dot{z} + \frac{i}{2}\left(\eta^i\dot{\bar{\eta}}_i-\dot{\eta}^i\bar{\eta}_i\right)+\frac{i}{2}\,\mu\left(v\dot{\bar{v}} + w\dot{\bar{w}} - \dot{v}\bar{v} - \dot{w}\bar{w}\right)\nn
	&& -\, \frac{i}{2}\,m\cos{2\lambda}\left(\dot{\bar{z}}\,\partial_{\bar z} K - \dot{z}\,\partial_{z} K\right)+ \frac{i\mu\left(\dot{\bar{z}}\,\partial_{\bar z}\bar{f} - \dot{z}\,\partial_z f\right)}{2{\left(f+\bar{f}\,\right)}}\left(v\bar{v}-w\bar{w}\right)\nn
    &&+\,\frac{i}{2}\left(\dot{\bar{z}}\,\partial_{\bar z} g -\dot{z}\,\partial_{z}g\right)g^{-1}\,\eta^k\bar{\eta}_k-\frac{m\cos{2\lambda}}{2}\,\eta^k\bar{\eta}_k-\frac{\mu\,\partial_z f\,\partial_{\bar z}\bar{f}}{\left(f+\bar{f}\,\right)^2}\left(v\bar{v}-w\bar{w}\right)g^{-1}\,\eta^k\bar{\eta}_k\nn
	&&-\,\frac{\mu\,v\bar{w}}{\left(f+\bar{f}\,\right)}\left[\frac{\partial_z\partial_z f}{2}-\frac{\partial_z f\,\partial_z f}{\left(f+\bar{f}\,\right)}-\frac{\partial_z f}{2}\,g^{-1}\,\partial_z g\right]g^{-1}\,\eta_i\eta^i\nn
	&&-\,\frac{\mu\,w\bar{v}}{\left(f+\bar{f}\,\right)}\left[\frac{\partial_{\bar z}\partial_{\bar z}\bar{f}}{2}-\frac{\partial_{\bar z}\bar{f}\,\partial_{\bar z}\bar{f}}{\left(f+\bar{f}\,\right)}-\frac{\partial_{\bar z} \bar{f}}{2}\,g^{-1}\,\partial_{\bar z} g\right]g^{-1}\,\bar{\eta}^j\bar{\eta}_j\nn
	&&+\,\frac{m\sin{2\lambda}}{4}\left[\left(\partial_z\partial_z K-g^{-1}\,\partial_z K\,\partial_z g\right)g^{-1}\,\eta_i\eta^i+\left(\partial_{\bar z}\partial_{\bar z} K-g^{-1}\,\partial_{\bar z}K\,\partial_{\bar z} g\right)g^{-1}\,\bar{\eta}^j\bar{\eta}_j\right]\nn
	&&-\,g^{-1}\left(\frac{\mu\,w\bar{v}\,\partial_z f}{\left(f+\bar{f}\,\right)}-\frac{m\sin{2\lambda}}{2}\,\partial_{z}K\right)\left(\frac{\mu\,v\bar{w}\,\partial_{\bar z}\bar{f}}{\left(f+\bar{f}\,\right)}-\frac{m\sin{2\lambda}}{2}\,\partial_{\bar z}K\right)\nn
	&&+\,\frac{1}{4}\left(\partial_{z}\partial_{\bar z}g - g^{-1}\,\partial_{z} g\,\partial_{\bar z}g\right)g^{-2}\,\eta_i\eta^i\,\bar{\eta}^j\bar{\eta}_j+\left[\mu\left(v\bar{v}+w\bar{w}\right)-c\right]{\cal A}\,.
\eea
Looking at the last term, one concludes that the U$(1)$ gauge field ${\cal A}$ plays the role of a Lagrange multiplier enforcing the constraint 
\bea
	\mu\left(v\bar{v}+w\bar{w}\right)-c=0.\label{spin}
\eea
\subsection{Hamiltonian and Noether charges}
Our next step is a formulation of a classical Hamiltonian mechanics. We perform Legendre transformation with the exclusion of the Lagrange multiplier ${\cal A}$\,.
As a result, we obtain the classical Hamiltonian
\bea
	{\cal H}&=&g^{-1}\left[p_z-\frac{i}{2}\,m\cos{2\lambda}\;\partial_z K+\frac{i\mu\,\partial_z f\left(v\bar{v}-w\bar{w}\right)}{2\left(f+\bar{f}\,\right)}+\frac{i}{2}\,g^{-1}\,\partial_z g\,\eta^{i}\bar{\eta}_i\right]\nn
	&&\times\left[p_{\bar z}+\frac{i}{2}\,m\cos{2\lambda}\;\partial_{\bar z}K-\frac{i\mu\,\partial_{\bar z}\bar{f}\left(v\bar{v}-w\bar{w}\right)}{2\left(f+\bar{f}\,\right)}-\frac{i}{2}\,g^{-1}\,\partial_{\bar z} g\,\eta^{j}\bar{\eta}_j\right]\nn
	&&+\,\frac{m\cos{2\lambda}}{2}\,\eta^k\bar{\eta}_k+\frac{\mu\,\partial_z f\,\partial_{\bar z}\bar{f}}{\left(f+\bar{f}\,\right)^2}\left(v\bar{v}-w\bar{w}\right)g^{-1}\,\eta^k\bar{\eta}_k\nn
	&&+\,\frac{\mu\,v\bar{w}}{\left(f+\bar{f}\,\right)}\left[\frac{\partial_z\partial_z f}{2}-\frac{\partial_z f\,\partial_z f}{\left(f+\bar{f}\,\right)}-\frac{\partial_z f}{2}\,g^{-1}\,\partial_z g\right]g^{-1}\,\eta_k\eta^k\nn
	&&+\,\frac{\mu\,w\bar{v}}{\left(f+\bar{f}\,\right)}\left[\frac{\partial_{\bar z}\partial_{\bar z}\bar{f}}{2}-\frac{\partial_{\bar z}\bar{f}\,\partial_{\bar z}\bar{f}}{\left(f+\bar{f}\,\right)}-\frac{\partial_{\bar z} \bar{f}}{2}\,g^{-1}\,\partial_{\bar z} g\right]g^{-1}\,\bar{\eta}^k\bar{\eta}_k\nn
	&&-\,\frac{m\sin{2\lambda}}{4}\left[\left(\partial_z\partial_z K-g^{-1}\,\partial_z K\,\partial_z g\right)g^{-1}\,\eta_i\eta^i +\left(\partial_{\bar z}\partial_{\bar z} K-g^{-1}\,\partial_{\bar z}K\,\partial_{\bar z} g\right)g^{-1}\,\bar{\eta}^j\bar{\eta}_j\right]\nn
	&&+\,g^{-1}\left(\frac{\mu\,v\bar{w}\,\partial_z f}{\left(f+\bar{f}\,\right)}-\frac{m\sin{2\lambda}}{2}\,\partial_{z}K\right)\left(\frac{\mu\,w\bar{v}\,\partial_{\bar z}\bar{f}}{\left(f+\bar{f}\,\right)}-\frac{m\sin{2\lambda}}{2}\,\partial_{\bar z}K\right)\nn
	&&-\,\frac{1}{4}\left(\partial_{z}\partial_{\bar z}g - g^{-1}\,\partial_{z} g\,\partial_{\bar z}g\right)g^{-2}\,\eta_i\eta^i\,\bar{\eta}^j\bar{\eta}_j\,,\label{H}
\eea
and the secondary constraint \eqref{spin}. Poisson (Dirac) brackets are imposed as
\bea
	&& \left\lbrace p_z\,, z\right\rbrace_{\rm P.B.} = -1,\qquad \left\lbrace p_{\bar z}\,, \bar{z}\right\rbrace_{\rm P.B.} = -1,\qquad\left\lbrace\eta^i,\bar{\eta}_j\right\rbrace_{\rm P.B.} = -i\delta^i_j\,,\nn
	&& \left\lbrace v, \bar{v}\right\rbrace_{\rm P.B.}=i\mu^{-1},\qquad \left\lbrace w, \bar{w}\right\rbrace_{\rm P.B.}=i\mu^{-1}.\label{brackets}
\eea
SU$(2|1)$ supercharges are 
\bea
	Q^i&=&\sqrt{2}\,e^{\frac{i}{2}mt}\,g^{-\frac{1}{2}}\Bigg\lbrace\cos{\lambda}\;\eta^i\left[p_z-\frac{i}{2}\,m\,\partial_z K+\frac{i\mu\,\partial_z f\left(v\bar{v}-w\bar{w}\right)}{2\left(f+\bar{f}\,\right)}+\frac{i}{2}\,g^{-1}\,\partial_z g\,\eta^{k}\bar{\eta}_k\right]\nn
	&&-\sin{\lambda}\;\bar{\eta}^i\left[p_{\bar z}-\frac{i}{2}\,m\,\partial_{\bar z}K-\frac{i\mu\,\partial_{\bar z}\bar{f}\left(v\bar{v}-w\bar{w}\right)}{2\left(f+\bar{f}\,\right)}-\frac{i}{2}\,g^{-1}\,\partial_{\bar z} g\,\eta^{j}\bar{\eta}_j\right]\Bigg\rbrace\nn
	&&-\,\frac{\sqrt{2}\,i\mu\,e^{\frac{i}{2}mt}\,g^{-\frac{1}{2}}}{\left(f+\bar{f}\,\right)}\left(\cos{\lambda}\;w\bar{v}\,\partial_{\bar z}\bar{f}\,\bar{\eta}^i-\sin{\lambda}\;v\bar{w}\,\partial_z f\,\eta^i\right),\nn
	\bar{Q}_j&=&\sqrt{2}\,e^{-\frac{i}{2}mt}\,g^{-\frac{1}{2}}\Bigg\lbrace\cos{\lambda}\;\bar{\eta}_j\left[p_{\bar z}+\frac{i}{2}\,m\,\partial_{\bar z}K-\frac{i\mu\,\partial_{\bar z}\bar{f}\left(v\bar{v}-w\bar{w}\right)}{2\left(f+\bar{f}\,\right)}-\frac{i}{2}\,g^{-1}\,\partial_{\bar z} g\,\eta^{k}\bar{\eta}_k\right]\nn
	&&+\sin{\lambda}\;\eta_j\left[p_z+\frac{i}{2}\,m\,\partial_z K+\frac{i\mu\,\partial_z f\left(v\bar{v}-w\bar{w}\right)}{2\left(f+\bar{f}\,\right)}+\frac{i}{2}\,g^{-1}\,\partial_z g\,\eta^{k}\bar{\eta}_k\right]\Bigg\rbrace\nn
	&&-\,\frac{\sqrt{2}\,i\mu\,e^{-\frac{i}{2}mt}\,g^{-\frac{1}{2}}}{\left(f+\bar{f}\,\right)}\left(\cos{\lambda}\;v\bar{w}\,\partial_z f\,\eta_j+\sin{\lambda}\;w\bar{v}\,\partial_{\bar z}\bar{f}\,\bar{\eta}_j\right).
\eea
With respect to the brackets \eqref{brackets}, they close on the superalgebra $su(2|1)$:
\bea
    &&\left\lbrace Q^{i}, \bar{Q}_{j}\right\rbrace_{\rm P.B.} = -\,i\left(2m I^i_j +2\delta^i_j {\cal H}\right),\qquad
    \left\lbrace I^i_j,  I^k_l\right\rbrace_{\rm P.B.}
    = -\,i\left(\delta^k_j I^i_l - \delta^i_l I^k_j\right),\nn
    &&\left\lbrace I^i_j, \bar{Q}_{l}\right\rbrace_{\rm P.B.} = -\,i\left(\frac{1}{2}\,\delta^i_j\bar{Q}_{l}-\delta^i_l\bar{Q}_{j}\right),\qquad \left\lbrace I^i_j, Q^{k}\right\rbrace_{\rm P.B.}
    = -\,i\left(\delta^k_j Q^{i} - \frac{1}{2}\,\delta^i_j Q^{k}\right),\nn
    &&\left\lbrace{\cal H}, Q^{i}\right\rbrace_{\rm P.B.}=\frac{i}{2}\,m\,Q^{i},\qquad \left\lbrace{\cal H}, \bar{Q}_{j}\right\rbrace_{\rm P.B.}=-\,\frac{i}{2}\,m\,\bar{Q}_{j}\,,
\eea
where the SU(2) generators are
\bea
	I^i_j=\eta^i\bar{\eta}_j-\frac{\delta^i_j}{2}\,\eta^k\bar{\eta}_k\,.\label{Iij}
\eea
It should be noted that the Hamiltonian \eqref{H} can always be redefined as a central charge when $\lambda=0$ (see the example in the next Section).

There is another SU$(2)$ subgroup that is generated by spin variables as
\bea
	S_3 = \frac{\mu}{2}\left(v\bar{v}-w\bar{w}\right),\qquad S_+ = \mu\,v\bar{w},\qquad S_- = \mu\,w\bar{v}.
\eea
Indeed, one can check these generators form the $su(2)$ algebra:
\bea
	\left\lbrace S_3, S_\pm\right\rbrace_{\rm P.B.}=\mp\,iS_\pm\,,\qquad \left\lbrace S_+, S_-\right\rbrace_{\rm P.B.}= -\,2i S_3\,.
\eea
Since the Hamiltonian \eqref{H} contains these SU$(2)$ generators, it commutes only with the quadratic Casimir operator
\bea
	C_{{\rm SU}(2)}=S_+S_- + \left(S_3\right)^2.
\eea
Taking into account \eqref{spin}, the Casimir operator is determined by
the constant
\bea
	C_{{\rm SU}(2)}=\frac{c^2}{4}\,.
\eea
Its quantum counterpart (up to the ordering ambiguity) is given by ($c \approx 2s$)
\bea
	C_{{\rm SU}(2)}=s\left(s+1\right),
\eea
where $s$ is a spin of the quantum system. 

Quantization of supersymmetric spinning models can be performed under the prescription of \cite{FIL2} (see also \cite{FIS, FILS}). Unlike the latter, the problem considered here has a more complex interaction with spin variables. There is also a peculiar fact that the Hamiltonian of supersymmetric ``K\"ahler oscillator'' models is associated with the U(1)-generator of \eqref{algebra1}. It has an impact on the energy spectrum, such that bosonic and fermionic states of a given SU$(2|1)$ representation have different energy values. Another feature of SU$(2|1)$ supersymmetry is that it can lead to spontaneously broken supersymmetry \cite{DHSS, FIS}.

\section{Model on SU(1,1)/U(1)}
Let us consider the model on the coset space SU$(1,1)$/U$(1)$, that is in fact a deformation of the model studied in \cite{BKKS}. It is given by the K\"ahler potential
\bea
	K=-\,\frac{1}{\gamma}\log{\left(1-\gamma\,z\bar{z}\right)}\,,\qquad \gamma >0,
\eea
that leads to the metric
\bea
	g=\frac{1}{\left(1-\gamma\,z\bar{z}\right)^2}\,.
\eea
To eliminate the terms $\sim\mu\,v\bar{w}\,\eta_k\eta^k$ in \eqref{H}, we choose the holomorphic function $f(z)$ as \footnote{In the limit $\gamma=0$ to the flat space, the coupling with the spin multiplet disappears since the holomorphic function \eqref{fSU11} becomes trivial: $f = 1$.}
\bea
	f=\frac{1+\sqrt{\gamma}\,z}{1-\sqrt{\gamma}\,z}\,.\label{fSU11}
\eea
To simplify the notations, we also perform a redefinition in the phase space that preserves the brackets \eqref{brackets}:
\bea
	&&v\rightarrow\left(1-\sqrt{\gamma}\,z\right)^{1/2}\left(1-\sqrt{\gamma}\,\bar{z}\right)^{-1/2}v,\qquad \bar{v}\rightarrow\left(1-\sqrt{\gamma}\,z\right)^{-1/2}\left(1-\sqrt{\gamma}\,\bar{z}\right)^{1/2}\bar{v},\nn
	&&\bar{w}\rightarrow\left(1-\sqrt{\gamma}\,z\right)^{1/2}\left(1-\sqrt{\gamma}\,\bar{z}\right)^{-1/2}\bar{w},\qquad w\rightarrow\left(1-\sqrt{\gamma}\,z\right)^{-1/2}\left(1-\sqrt{\gamma}\,\bar{z}\right)^{1/2}w,\nn
	&&p_z \rightarrow p_z - \frac{i\,\sqrt{\gamma}\,\mu\left(v\bar{v}-w\bar{w}\right)}{2\left(1-\sqrt{\gamma}\,z\right)}\,,\qquad
	p_{\bar z} \rightarrow p_{\bar z} + \frac{i\,\sqrt{\gamma}\,\mu\left(v\bar{v}-w\bar{w}\right)}{2\left(1-\sqrt{\gamma}\,\bar{z}\right)}\,.
\eea
Thereby, the Hamiltonian reads
\bea
	{\cal H}&=&\left(1-\gamma\,z\bar{z}\right)^2\left[\bar{\pi}-\frac{i\left(1-\cos{2\lambda}\right)mz}{2\left(1-\gamma\,z\bar{z}\right)}\right]\left[\pi+\frac{i\left(1-\cos{2\lambda}\right)m\bar{z}}{2\left(1-\gamma\,z\bar{z}\right)}\right]+\frac{m\cos{2\lambda}}{2}\,\eta^k\bar{\eta}_k\nn
	&&+\,\gamma\,\mu\left(v\bar{v}-w\bar{w}\right)\eta^k\bar{\eta}_k+\frac{\gamma\,m\sin{2\lambda}}{2}\left(\bar{z}^2\,\eta_k\eta^k+z^2\,\bar{\eta}^k\bar{\eta}_k\right)-\frac{\gamma}{2}\,\eta_i\eta^i\,\bar{\eta}^j\bar{\eta}_j\nn
	&&+\left(\sqrt{\gamma}\,\mu\,v\bar{w}-\frac{m\sin{2\lambda\;\bar{z}}}{2}\right)\left(\sqrt{\gamma}\,\mu\,w\bar{v}-\frac{m\sin{2\lambda}\;z}{2}\right),\label{HSU11}
\eea
where
\bea
	&&\pi = p_z - \frac{i\bar{z}}{1-\gamma\,z\bar{z}}\left[\frac{m}{2} +\frac{\gamma\,\mu}{2}\left(v\bar{v}-w\bar{w}\right)-\gamma\,\eta^{i}\bar{\eta}_i\right],\nn
	&&\bar{\pi} = p_{\bar z}+\frac{iz}{1-\gamma\,z\bar{z}}\left[\frac{m}{2} +\frac{\gamma\,\mu}{2}\left(v\bar{v}-w\bar{w}\right)-\gamma\,\eta^{i}\bar{\eta}_i\right].
\eea
Supercharges are written in the form
\bea
	Q^i&=&\sqrt{2}\,e^{\frac{i}{2}mt}\left(1-\gamma\,z\bar{z}\right)\left[\cos{\lambda}\;\eta^i\,\pi-\sin{\lambda}\;\bar{\eta}^i\left(\bar{\pi}-\frac{imz}{1-\gamma\,z\bar{z}}\right)\right]\nn
	&&-\,\sqrt{2\gamma}\,e^{\frac{i}{2}mt}\,i\mu\left(\cos{\lambda}\;w\bar{v}\,\bar{\eta}^i-\sin{\lambda}\;v\bar{w}\,\eta^i\right),\nn
	\bar{Q}_j&=&\sqrt{2}\,e^{-\frac{i}{2}mt}\left(1-\gamma\,z\bar{z}\right)\left[\cos{\lambda}\;\bar{\eta}_j\,\bar{\pi}+\sin{\lambda}\;\eta_j\left(\pi+\frac{im\bar{z}}{1-\gamma\,z\bar{z}}\right)\right]\nn
	&&-\,\sqrt{2\gamma}\,e^{-\frac{i}{2}mt}\,i\mu\left(\cos{\lambda}\;v\bar{w}\,\eta_j+\sin{\lambda}\;w\bar{v}\,\bar{\eta}_j\right).
\eea
SU$(2)$ generators are given by the same expression \eqref{Iij}.

\paragraph{Limit $\lambda=0$.} In distinct from \cite{BKKS}, the Hamiltonian \eqref{HSU11} contains terms ($\sim m\sin{2\lambda}$) breaking the target space symmetry SU$(1,1)$. For instance, the corresponding bosonic Hamiltonian is characterized by a term such as the oscillator term $m^2\sin^2{2\lambda}\;z\bar{z}/4$\,. We can define only the remaining U$(1)$ generator $J_3$\,,
\bea
	J_3 = i\left(\bar{z}p_{\bar z}-zp_z\right)+\frac{\mu}{2}\left(v\bar{v}-w\bar{w}\right)-\eta^k\bar{\eta}_k\,,
\eea
that commutes with all SU$(2|1)$ generators. Thus, to restore the symmetry SU$(1,1)$ we take the limit $\lambda=0$. In this limit, we can introduce an external U$(1)$ generator $F$ given by the expression
\bea
	F=\frac{1}{2}\left[\eta^k\bar{\eta}_k-\mu\left(v\bar{v}-w\bar{w}\right)\right],
\eea
that rotates the relevant supercharges for $\lambda=0$ as
\bea
    \left\lbrace F, Q^{i}|_{\lambda=0}\right\rbrace_{\rm P.B.}=-\,\frac{i}{2}\,Q^{i}|_{\lambda=0}\,,\qquad \left\lbrace F, \bar{Q}^{j}|_{\lambda=0}\right\rbrace_{\rm P.B.}=\frac{i}{2}\,\bar{Q}^{j}|_{\lambda=0}\,.
\eea
Hence, we can define a central charge generator of the extended superalgebra $su(2|1) {+\!\!\!\!\!\!\supset}\, u(1)$ as
\bea
	H={\cal H}|_{\lambda=0} + mF.\label{CC}
\eea
and take it as a new Hamiltonian of the system, while the generator $F$ becomes an internal U$(1)$ generator of the extended superalgebra. 

Deformed SU$(1,1)$ symmetry generators are written as
\bea
	&&J_3 = i\left(\bar{z}p_{\bar z}-zp_z\right)+\frac{\mu}{2}\left(v\bar{v}-w\bar{w}\right)-\eta^k\bar{\eta}_k\,,\nn
	&&J_+ = p_{\bar z} - \gamma\,z^2 p_z - iz\left[\frac{m}{2} +\frac{\gamma\,\mu}{2}\left(v\bar{v}-w\bar{w}\right)-\gamma\,\eta^{k}\bar{\eta}_k\right],\nn
	&&J_- = p_z - \gamma\,\bar{z}^2 p_{\bar z}+i\bar{z}\left[\frac{m}{2} +\frac{\gamma\,\mu}{2}\left(v\bar{v}-w\bar{w}\right)-\gamma\,\eta^{k}\bar{\eta}_k\right].
\eea
Indeed, they commute with all SU$(2|1)$ generators including $F$, and form the central extension of the $su(1,1)$ algebra:
\bea
	\left\lbrace J_3, J_\pm\right\rbrace_{\rm P.B.}=\pm\,iJ_\pm\,,\qquad \left\lbrace J_+, J_-\right\rbrace_{\rm P.B.}= -\,2i\gamma J_3-im,
\eea
where the deformation parameter $m$ is a central charge.
Defining its Casimir operator
\bea
	C_{{\rm SU}(1,1)}=J_+J_- - \gamma\left(J_3+\frac{m}{2\gamma}\right)^2,
\eea
we obtain the following expression for the Hamiltonian \eqref{CC}:
\bea
	H=\gamma\,C_{{\rm SU}(2)}+C_{{\rm SU}(1,1)}+\frac{m^2}{4\gamma}\,.
\eea
In the limit $m=0$, we reproduce the model studied in \cite{BKKS}.

In the limit $\gamma=0$, the generators $J_{\pm}$ become magnetic translation generators forming the Heisenberg algebra with the external U$(1)$ automorphism group generator $J_3$:
\bea
	\left\lbrace J, J_\pm\right\rbrace_{\rm P.B.}=\pm\,iJ_\pm\,,\qquad \left\lbrace J_+, J_-\right\rbrace_{\rm P.B.}= -\,im,\qquad \left\lbrace J_3, J_\pm\right\rbrace_{\rm P.B.}=\pm\,iJ_\pm\,.
\eea 
Spin variables drop out of the model leaving only the dynamical multiplet ${\bf (2,4,2)}$ \cite{DSQM}.

\section{Conclusions}

In this paper, we proposed new models of single-particle SU$(2|1)$ supersymmetric mechanics with the use of dynamical, semi-dynamical and gauge multiplets. As an alternative of harmonic superspace, we exploited the generalized chiral superspace \eqref{left}. The generalized chiral multiplet ${\bf (2,4,2)}$ was introduced in \cite{SKO}. For the mirror multiplet ${\bf (4,4,0)}$, we defined the gauge invariant constraints \eqref{1} and \eqref{2}. As turned out, the gauge superfield $X$, bonded with these constraints, satisfies \eqref{143c} and describes the mirror multiplet ${\bf (1,4,3)}$. As an example, we considered in details the deformed model on the pseudo-sphere SU$(1,1)$/U$(1)$. One could consider as an example the quantum model on a complex plane ($g=1$), but even in the limit $\lambda=0$ the Hamiltonian \eqref{H} seems to have a non-trivial potential.

The most natural question is whether it is possible to consider superconformal models. During our study of superconformal models of the trigonometric type \cite{IST}, we suggested a simple criteria for  construction of superconformal models. With this experience we tried to obtain superconformal symmetry for considered models, but it is impossible at least for a single-particle model. For example, if we take two chiral dynamical multiplets then suitable superconformal models can possibly be obtained. This may be a good challenge for future research. 

It would be interesting to study the mirror counterpart of the multiplet ${\bf (1,4,3)}$ \cite{DSQM}. Another interesting problem is the application of the gauging procedure \cite{FI} to the deformed multiplets ${\bf (4,4,0)}$ \cite{DHSS}. It can give rise to general actions of the ordinary and mirror multiplets ${\bf (3,4,1)}$ \footnote{SU$(2|1)$ constraints for the ordinary one were solved in Appendix C of \cite{ILS}.}.

\section*{Acknowledgments}
\noindent The author is indebted to Evgeny Ivanov for interest and valuable comments. The author also thanks Sergey Fedoruk and Armen Nersessian for useful discussions. This research was supported by the Russian Science Foundation Grant No 16-12-10306.

\end{document}